# 一种应用于空间碎片演化模型的碰撞概率算法

王晓伟[1,2,3]，刘 静[1,2]，崔双星[1,2]

(1. 中国科学院国家天文台，北京，100101；2. 国家航天局空间碎片监测与应用中心，北京，100101；3. 中国科学院大学，北京，100049)

**摘 要**：对碰撞概率算法Cube模型参数影响空间碎片演化模型的仿真结果问题进行了深入分析与研究，对原Cube算法做了改进，提出了I-Cube模型。经过多次蒙特卡洛仿真结果验证，I-Cube模型对演化过程中空间碎片碰撞概率的计算更为准确合理，空间碎片长期演化模型的结果不再依赖于自身碰撞概率算法的相关参数，提高了空间碎片长期演化模型的稳定性与可信度。

**关键词**：空间碎片；长期演化；碰撞概率；Cube模型

**中图分类号**：P139　　**文献标识码**：A　　**文章编号**：

## A Collision Probability Estimation Algorithm Used in the Space Debris Evolutionary Model

WANG Xiao-wei[1,2,3], LIU Jing[1,2], CUI Shuang-xing[1,2]

（1. National Astronomical Observatories, Chinese Academy of Sciences, Beijing, 100101;
2. Space Debris Observation and Data Application Center, CNSA, Beijing, 100101;
3. University of Chinese Academy of Sciences, Beijing, 100049）

**Abstract**： An in-depth analysis is performed on the problem that one parameter of the Cube model can affects the final simulation results of space debris long-term evolution model, which weakens the representativeness of the space debris evolution model. We made some improvements and proposed an Improved-Cube (I-Cube) model. By multiple Monte Carlo simulations, it is indicated that the I-Cube model offered a more accurate and more reasonable option for collision probability estimation in the space debris evolution process. The simulation results of space debris long-term evolution model are no longer sensitive to the collision probability estimation model parameters, thus improved the reliability of space debris long-term evolution model.

**Key words**： Space debris; Long-term evolution; Collision probability; Cube model

## 0 引 言

近年来，空间碎片数量激增使得在轨航天器的运行安全受到极大威胁，Kessler雪崩效应再次被提及[1]。空间碎片数量不断增长引发了国际上关于空间碎片环境长期稳定性的研究，空间碎片环境长期演化模型成为该领域的国际研究热点之一[2-9]。

空间碎片环境长期演化模型可以预测未来几十年至上百年的空间碎片环境演化，为调整太空发展战略或制定相关空间政策提供技术支持，以保证未来太空活动可持续发展。空间碎片演化模型能够模拟空间碎片的主要增长机制和减少机制，如未来的航天器发射、在轨碰撞或爆炸解体、自然陨落、任务后处置等，通常由轨道预报模型、未来发射模型、碰撞概率评估模型、解体模型、任务后处置模型等几个子模型组成。由于在轨碰撞解体是空间碎片在未来演化中的重要增长来源，对碰撞概率的评估会影响空间碎片在轨碰撞事件的预测，因此碰撞概率评估模型是空间碎片演化模型中较为关键的环节之一。

传统的瞬时碰撞概率计算方法多是基于空间物体的位置椭球误差[10-11]，然而空间碎片演化模型预测的时间较长（几十年甚至几百年），空间物体轨道预报的精度有限，且高精度的位置误差对于长期演化并不具有实际意义，因此传统的瞬时碰撞概率计算方法并不适用。要预测空间碎片在未来长期演化中发生碰撞解体事件的可能性，需要一种适用于长期轨道演化系统的碰撞概率估计方法。

Cube模型是由美国国家宇航局（National Aeronautics and Space Administration, NASA）提出的一种快速成对算法，适用于任何轨道演化系统。Cube算法通过对整个演化系统进行时间均匀采样，能够利用演化过程中不断更新的轨道根数来评估空间物体之间的碰撞概率[12-13]。Cube模型算法具有原理简单、配对快速的优点。Cube模型算法提出后，被多个国家的空间碎片环境长期演化模型所采纳。但在

2014 年第 32 届机构间空间碎片协调委员会（Inter-Agency Space Debris Coordination Committee, IADC）全体会议上，法国宇航局（Centre National d'Etudes Spatiales, CNES）指出，利用其演化模型 MEDEE 模拟低地球轨道（Low Earth Orbit, LEO）空间环境的未来演化，当 Cube 模型的立方体尺寸取值不同时，空间环境演化结果会有较大差异，如图 1 所示，其中实线表示 40 次蒙特卡洛运行的平均结果，虚线表示 1σ 的标准差；表 1 为 MEDEE 模型预测的 200 年后的 LEO 空间碎片数量相比于初始碎片数量的变化百分比，及其 1σ 标准差[14-15]。但从物理规律而言，空间碎片在演化过程中的碰撞概率不会因碰撞概率评估算法不同而改变，因此若 Cube 模型因其参数设置不同而对未来空间环境演化结果带来显著影响，将直接导致空间碎片演化模型的可信度降低。

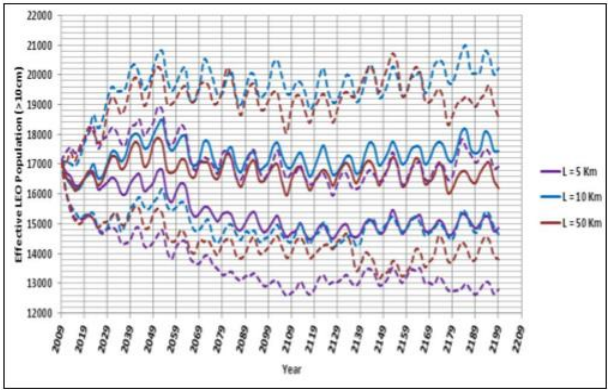

图 1 立方体大小不同时 MEDEE 模型的运行结果
Fig.1 Evolution results of MEDEE with different cube sizes

表 1 200 年后空间碎片数量的变化百分比及其 1σ 标准差
Table 1 The percentage of variation of the number of objects after 200 years, together with 1-σ dispersion.

| Cube size | % of Variation wrt Initial population after 200 years [mean +/- 1 σ Dispersion] |
|---|---|
| L=5 Km | -13% +/- 12% |
| L=10 Km | 2% +/- 16% |
| L=50 Km | -5.3% +/- 14% |

本文对此问题进行了深入分析，研究了 Cube 模型算法及其在空间碎片演化模型中的应用问题，并在 Cube 模型算法的基础上做了改进，提出了 I-Cube 模型。经过多次蒙特卡洛模拟运行验证，使用改进后的 I-Cube 算法，空间碎片演化模型的演化结果不再受碰撞概率算法参数的影响，显著提高了演化模型的稳定性与可信度。

本文所使用的 SOLEM 模型（Space Objects Long-term Evolution Model）是我国自主建立的空间碎片长期演化模型[16-18]。SOLEM 模型作为中国国家航天局（China National Space Administration, CNSA）的代表参与了 IADC 组织的多项国际联合研究，并取得了与国际上其他演化模型较为一致的结果[19]。

本文第 1 部分对 Cube 模型算法及其在空间碎片演化模型中的应用做了进一步说明与分析，第 2 部分介绍改进的碰撞概率模型 I-Cube，第 3 部分对比分析了采用 Cube 模型算法与 I-Cube 算法的演化结果，第 4 部分进行了总结。

## 1 碰撞概率模型及其对演化的影响

碰撞概率的计算直接影响空间碎片碰撞解体次数的预测。碰撞解体是未来空间碎片数量增长的重要来源，因而对空间碎片间的碰撞概率估计成为影响未来空间环境演化仿真结果的关键环节之一。

### 1.1 碰撞概率算法 Cube 模型

Cube 模型是由 NASA 提出的适用于任何轨道演化系统的碰撞概率评估算法[12-13]。Cube 算法通过对整个演化系统进行时间均匀采样的方法来计算空间碎片的碰撞概率。数学上，物体 $i$ 和 $j$ 之间在很长一段时间内（从 $t_{begin}$ 到 $t_{end}$）的总碰撞次数可表达为：

$$N_{tot} = \int_{t_{begin}}^{t_{end}} P_{i,j}(t)\mathrm{d}t = \int_0^L \int_{t_s}^{t_{s+1}} P_{i,j}(t)\mathrm{d}t\mathrm{d}s \quad (1)$$

式中：$P_{i,j}$ 是碰撞率，$L$ 是 $t_{begin}$ 和 $t_{end}$ 之间的时间间隔数，$t_s$ 和 $t_{s+1}$ 代表第 $s$ 个时间间隔的起止时刻。如果时间间隔 $t_{s+1} - t_s$ 足够短，两个空间碎片之间的碰撞特征变化不大，那么可认为 $P_{i,j}$ 在这段时间内是常数，则上述积分表达式可写为

$$N_{tot} = \int_0^L [t_{s+1} - t_s] \times P_{i,j}(s)\mathrm{d}s \quad (2)$$

在每次采样时刻，建立地心笛卡尔坐标系，将三维近地空间划分成多个边长为 $h$ 的小立方体，每个空间物体的位置、速度根据该时刻的轨道根数计算。当两个空间物体处于同一个立方体时，碰撞率 $P_{ij}$ 由下式计算

$$P_{ij} = s_i s_j A_c V_{imp} \mathrm{d}U \quad (3)$$

式中：$s_i$ 和 $s_j$ 表示目标 $i$ 和 $j$ 在该立方体内的空间

密度，$A_c$ 表示两目标的碰撞截面，$V_{imp}$ 表示两个目标的碰撞速度，$dU$ 表示该立方体的体积。对 $P_{ij}$ 进行时间积分后得到碰撞概率

$$p_{ij} = P_{ij} \cdot [t_{s+1} - t_s] = s_i s_j A_c V_{imp} dU dt \quad (4)$$

计算出 $p_{ij}$ 后，应用蒙特卡罗方法生成一个随机数 $R_i$ 与 $p_{ij}$ 进行比较，以此来确定一次碰撞是否发生。若一个立方体内同时有两个或多个物体，计算其两两之间的碰撞概率，并确定碰撞是否发生；而若一个立方体内只有一个空间物体，则不再考虑该物体与其他物体的碰撞可能性。对于 $N$ 体系统，该方法能够快速有效地找出碰撞配对物体，计算时间与 $N$ 成正比，而非 $N^2$。

根据标准统计采样方法，采样次数越多，即采样时间间隔越短越好，但在空间碎片长期演化模型中，需要考虑整个模型的运行速度，不能在碰撞概率计算上花费过多的时间。在美国的空间碎片长期演化模型 LEGEND 中，$dt$ 默认值为 5 天[13]。

而关于采样时刻所划分的立方体大小，Cube 模型认为在平均轨道半长轴的 1% 及以内都是合理的[13]。将其应用在空间碎片演化模型中时，主要研究对象是 LEO 轨道的碎片，轨道半长轴在 6578 km ~ 8578 km 之间，故依据 Cube 模型，其所划分的立方体尺寸在 65 km 及以下时都是合理的。美国空间碎片长期演化模型 LEGEND 中，$h$ 取值为 10 km[13]。

**1.2 Cube 算法存在的问题**

依据 Cube 模型，上述碰撞概率算法式（4）相当于在一个微观尺度上（一个立方体内）应用气体动力学理论。然而，根据式（4）计算出的 $p_{ij}$ 并非真正的碰撞概率，而是两个物体在所划分的立方体 $dU$ 内在积分时间 $dt$ 内的平均碰撞次数。在积分时间并非趋于 0 的情况下，对于个别截面积超大的碎片，用该式计算出的碰撞概率存在 $p_{ij}>1$ 的情况。实际上，在 $p_{ij} \leq 0.2$ 的情况下，可将其近似看作碰撞概率，其误差小于 10%，而当 $p_{ij}>0.2$ 时，该近似不再成立，必须使用碰撞概率的严格表达式（见式（5））[20]。

此外，依据 Cube 模型假设，当划分的立方体足够大时，所有空间物体都处在同一个立方体内，因而需要计算所有空间物体两两之间的碰撞概率；反之，当划分的立方体足够小时，没有任何两个及以上的空间物体能同时处在一个立方体内，从而不需要计算任何物体之间的碰撞概率。因此，Cube 算法的假设，只有位于同一个立方体内的空间物体间的碰撞概率非零，不符合物理真实。即使给出了一个合理范围内的参数 $h$，如 10 km，对于同一个立方体内的两个空间物体，其距离最大值为立方体对角线 $\sqrt{3}h$。实际上，在相邻立方体中，任何两个距离小于立方体对角线 $\sqrt{3}h$ 的空间物体都有可能位于某个方向上的同一立方体内，然而 Cube 算法并未考虑其碰撞可能性。在采样时刻，决定是否考虑两个空间物体间的碰撞可能性完全取决于立方体的划分方式。因此，依据 Cube 模型假设，只考虑存在于同一个立方体中的两个物体之间的碰撞概率，势必导致演化仿真结果依赖于相关参数。图 2 为 Cube 参数 $h$ 取值分别为 5 km、8 km、10 km、30 km、50 km 时，利用 SOLEM 模型模拟的未来空间碎片的演化结果。

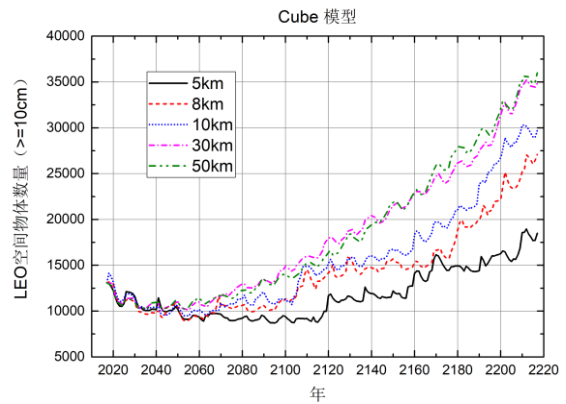

图 2 立方体大小 $h$ 对空间碎片演化预测结果的影响

Fig.2 Evolution results of SOLEM model with different cube size $h$

## 2 改进的碰撞概率算法 I-Cube 模型

针对上述问题我们在 Cube 模型的基础上做了改进，称为 Improved-Cube（I-Cube）模型。I-Cube 模型中假设只要与目标物体的距离满足阈值要求的碎片都具有与目标物体碰撞的可能性。其适用条件与原 Cube 算法一致。在寻找碰撞配对物体时，仍沿用 Cube 算法在采样时刻对空间划分立方体的思路，但在考虑同一立方体内两个及以上物体之间的碰撞概率外，也考虑与周边相邻立方体中距离小于立方

体对角线 $\sqrt{3}h$ 的空间物体之间碰撞概率。这一点与英国航天局（United Kingdom Space Agency, UKSA）的空间碎片长期演化模型 DAMAGE 考虑一致[18]。

寻找碰撞配对物体的空间范围增大后，应用气体动力学理论的空间范围也随之变大，因此 $dU$ 表示的不再是立方体体积，而是以立方体对角线 $\sqrt{3}h$ 为半径的球体积（如图 3 所示）。

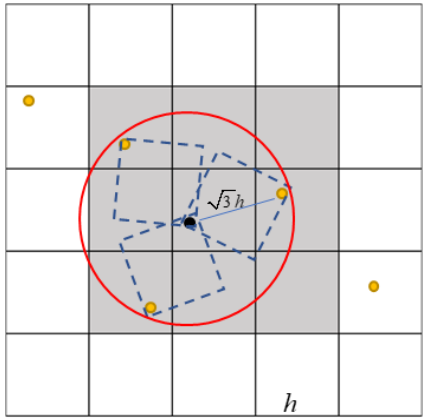

图 3 考虑相邻立方体中两个物体的碰撞概率的二维示意图
Fig.3 Two-dimensional representation for considering possible collisions between debris residing in neighbouring cubes

I-Cube 算法的具体实现步骤如下：
1) 确定对演化系统进行时间均匀采样的间隔 $dt$。在 SOLEM 模型中，时间采样间隔 $dt$ 为 5 天。
2) 对近地空间划分立方体。在采样时刻，建立地心笛卡尔坐标系，将地球周围的近地空间划分为一个个边长为 $h$ 的立方体，并对每个立方体按 $h$ 的倍数进行编号。
3) 计算空间物体所在的位置。在采样时刻，根据每个空间物体（序号为 $i$）更新后的轨道根数计算其所在的位置，并记录其所处的立方体编号。
4) 对所有空间物体进行快速配对。在采样时刻，对所有空间物体所在的立方体编号进行比对，找出处在相邻的或者相同的立方体内的目标，保留距离在阈值 $d_c$（如 $\sqrt{3}h$）之内的目标作为配对目标。
5) 计算配对目标之间的碰撞概率 $p_{ij}$。
6) 利用蒙特卡罗方法判断碰撞是否实际发生。生成 0 到 1 之间的随机数 $R_i$，若 $R_i < p_{ij}$，则碰撞发生，否则不发生。

步骤 5) 中碰撞概率 $p_{ij}$ 的具体算法如下：

根据气体动力学，在体元 $dU$ 内，在 $dt$ 时间内，空间物体 $i$ 和 $j$ 的平均碰撞数为

$$c = S_i S_j V_{imp} A_c dU dt \quad (5)$$

式中：$S_i$、$S_j$ 是两个空间物体在体元 $dU$ 内的分布密度，$V_{imp}$ 是两个空间物体的相对碰撞速度，$A_c$ 为两个碎片的碰撞截面。碰撞次数服从泊松统计，发生 0 次碰撞的概率为 $P_{i=n=0} = \frac{c^n}{n!}\exp(-c) = \exp(-c)$，故发生碰撞的概率为

$$p_{ij} = 1 - \exp(-c) \quad (6)$$

本方法中考虑的配对目标是欧氏距离在 $d_c$ 之内的空间物体，$dU$ 不再表示立方体的体积，而是以 $d_c$ 为半径的球体积，即 $dU = \frac{4\pi}{3}d_c^3 = 4\pi \cdot \sqrt{3} \cdot h^3$。综上可计算得到空间物体 $i$ 和 $j$ 之间的碰撞概率 $p_{ij}$。

与 Cube 算法相比，I-Cube 算法采用了严格表达式来计算碰撞概率 $p_{ij}$，此外增加了处于相邻立方体内的空间物体间碰撞可能性的计算，寻找碰撞配对物体的复杂度增加，因而计算所用的时间也相应增加，基本相当于原 Cube 算法计算时间的 3 倍。但从整个演化模型运行上看，这个计算速度仍在可接受范围内。

## 3 仿真分析

据目前公开的文献看，只有 CNES 利用其 MEDEE 模型对原 Cube 算法参数的影响做过相关研究[14-15]，此外，UKSA 的 DAMAGE 模型采用 Cube 算法时对其做了部分改进[21]。但由于各国的空间碎片长期演化模型在各个模块算法的具体技术实现方式并未公开，MEDEE 模型、DAMAGE 模型以及国际上其他长期演化模型在 Cube 算法的应用或改进方面，以及太阳活动、大气模型等关键影响因素的

假设与参数设置方面均与 SOLEM 模型不尽相同，且在已公开的文献中，MEDEE 和 DAMAGE 模型与 SOLEM 模型所使用的初始空间碎片数据也不同，故无法就 Cube 算法改进前后其参数对空间环境演化仿真的影响与 SOLEM 模型形成横向对比。鉴于 SOLEM 模型在国际联合研究中的出色表现，并出于避免不同演化模型因模型本身差异带来影响的考虑，本文仅利用 SOLEM 模型在保证其他关键影响因素保持不变的前提下，对 Cube 算法改进前后改变参数 $h$ 对演化结果的影响进行研究。

SOLEM 演化模型中将太阳辐射流量和地磁活动设为平均值，即 F10.7=130 sfu，Ap=9。轨道预报算法为简化的半分析法，考虑地心引力、地球非球形引力摄动、大气阻力摄动、第三体引力摄动和光压摄动，大气密度模型采用 NRLMSIS00 模型。碰撞解体模型采用 NASA 标准解体模型。以 2017 年 9 月 1 日的空间碎片环境数据作为初始输入，预报 200 年至 2217 年 9 月 1 日。以 2009.09.01 至 2017.08.31 之间 8 年的发射情况作为发射模型在未来 200 年内不断循环，航天器任务寿命假设为 8 年，不考虑轨道维持和人工避碰措施，任务后轨道处置率假设为 30%。

SOLEM 模型在上述假设条件均不变的情况下，碰撞概率评估模型分别采用 Cube 算法和 I-Cube 算法，仅改变参数 $h$ 的大小，研究其对未来 200 年空间碎片环境演化的影响。本文算例中参数 $h$ 取值分别为 5 km、8 km、10 km、30 km、50 km。每个算例均运行 50 次蒙特卡洛模拟并求平均。仿真结果如图 4-7 所示，其结果量化如表 2 所示。

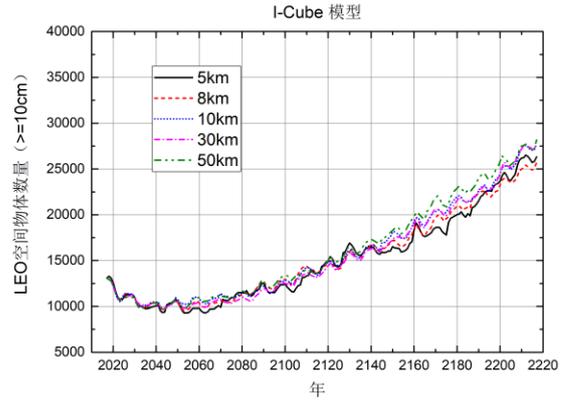

图 5 采用 I-Cube 算法，未来空间碎片数量的演化
Fig.5 Predictions of future space environment evolution using I-Cube method

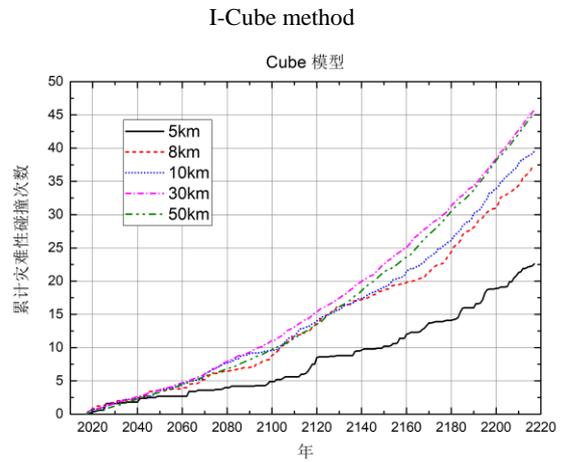

a 灾难性碰撞次数

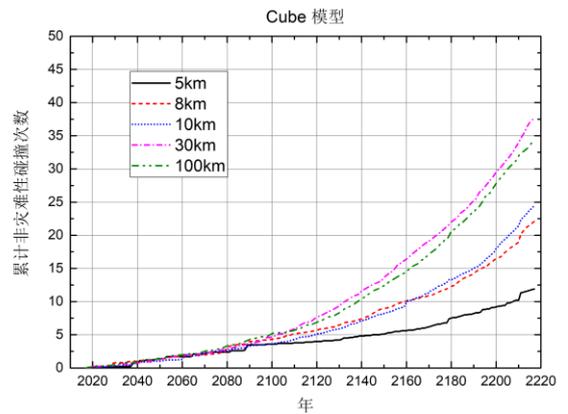

b 非灾难性碰撞次数

图 6 采用 Cube 算法，未来 200 年空间碎片碰撞次数的预测
Fig.6 Predictions of cumulative number of space debris collisions in future 200 years using Cube method

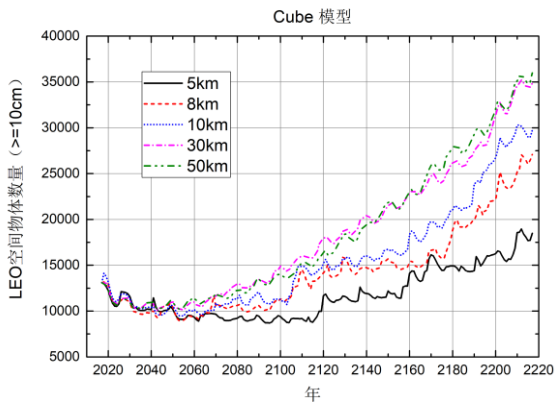

图 4 采用 Cube 算法，未来空间碎片数量的演化
Fig.4 Predictions of future space environment evolution using Cube method

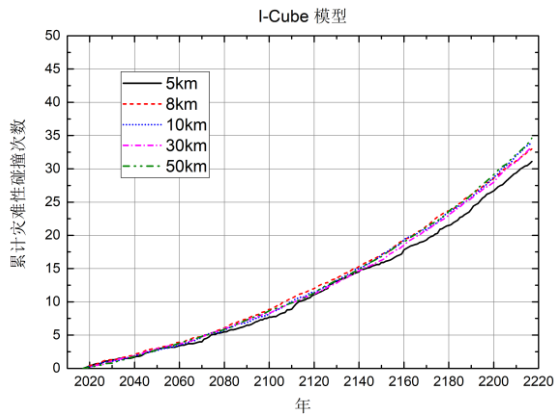

a 灾难性碰撞次数

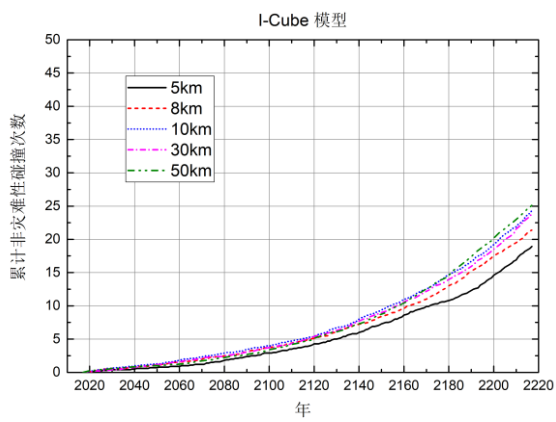

b 非灾难性碰撞次数

图 7 采用 I-Cube 算法，未来 200 年空间碎片碰撞次数的预测

Fig.7 Predictions of cumulative number of space debris collisions in future 200 years using I-Cube method

图 4-7 以及表 2 所示的仿真结果表明，演化模型在其他如太阳辐射等关键影响因素保持不变的情况下，采用原 Cube 算法，改变所划分立方体的大小 $h$ 会严重影响未来空间碎片环境的演化结果。就算例中 $h$ 的取值范围（5 km – 50 km），不同的取值造成 200 年后空间碎片数量演化的预测值相差约 17500 个，为 2017 年初始碎片数量的 134%；200 年累计灾难性碰撞次数预测相差 23 次，非灾难性碰撞次数预测相差约 27 次。由于 $h$ 的取值皆在"平均轨道半长轴的 1% 及以内"的合理范围，因而无法判断哪种演化结果可信。

而采用 I-Cube 算法后，改变参数 $h$ 大小，未来空间碎片环境的演化结果不再受其显著影响。就算例中 $h$ 的取值范围（5 km - 50 km），不同的取值造成 200 年后空间碎片数量演化的预测值相差仅为 2017 年初始碎片数量的 18%，200 年累计灾难性碰撞次数预测相差不到 4 次，非灾难性碰撞次数预测相差约 6 次。

由此可见，采用 I-Cube 算法后，未来空间碎片的演化结果高度一致，碰撞解体事件的预测不再受参数 $h$ 的显著影响。相比于采用 Cube 算法，I-Cube 算法对碰撞概率的计算更为准确合理，进而使空间碎片演化模型的稳定性与可信度显著提升。

表2 采用Cube算法与I-Cube算法，改变参数 $h$ 对未来空间环境演化的影响对比（红色字体表明最大和最小）

Table2 Impacts of varying parameter values of $h$ on the predictions of future space debris environment evolution using Cube algorithm compared with those using I-Cube algorithm

| 立方体大小 | Cube 算法 | | | I-Cube 算法 | | |
| --- | --- | --- | --- | --- | --- | --- |
| | 200 年后碎片增长百分比 | 灾难性碰撞次数 | 非灾难性碰撞次数 | 200 年后碎片增长百分比 | 灾难性碰撞次数 | 非灾难性碰撞次数 |
| $h$=5 km | 41% | 22.6 | 11.9 | 101% | 31.1 | 18.9 |
| $h$=8 km | 107% | 37.4 | 22.1 | 97% | 33.0 | 21.4 |
| $h$=10 km | 127% | 39.5 | 24.5 | 112% | 34.0 | 24.3 |
| $h$=30 km | 166% | 45.7 | 37.8 | 112% | 33.2 | 23.8 |
| $h$=50 km | 175% | 45.3 | 38.6 | 115% | 34.6 | 25.1 |
| 不确定度 | 134% | 23.1 | 26.7 | 18% | 3.5 | 6.2 |

## 4 结 论

I-Cube 算法采用了严格表达式来计算空间物体间的碰撞概率，排除了原 Cube 算法中可能出现"碰撞概率"大于 1 的隐患。

I-Cube 算法与 Cube 算法的模型假设不同。

I-Cube 算法假设只要两个物体间的距离在给定阈值范围内都存在碰撞可能性，而 Cube 算法假设只有位于同一个立方体内的物体之间存在碰撞可能性。相比之下，I-Cube 算法的模型假设更符合物理真实。

I-Cube 算法保留原 Cube 算法寻找碰撞配对物体的思路，仍通过在采样时刻对近地空间划分立方体的方法来快速寻找潜在的碰撞对，但寻找潜在碰撞对的范围由原来的同一立方体扩展至相邻的立方体。由于增加了寻找碰撞配对物体的复杂度，I-Cube 算法牺牲了一部分计算时间，但从空间碎片演化模型的整体运行上看，计算速度仍可接受。

相比于 Cube 算法，I-Cube 算法提升了碰撞概率计算方法的准确度与合理性，使空间碎片环境长期演化模型的演化结果不再受自身碰撞概率算法参数的影响，从而提高了空间碎片长期演化模型的稳定性与可信度。

## 参 考 文 献

作者简介：

**王晓伟**（1987-），女，博士生，主要从事空间碎片环境演化建模相关研究

通信地址：北京市朝阳区大屯路甲 20 号国家天文台（100101）

电话：（010）64807861

E-mail:xwwang@bao.ac.cn